\documentclass[conference]{IEEEtran}
\usepackage[utf8]{inputenc}
\usepackage{amsmath,amssymb}
\usepackage{graphicx}
\usepackage{booktabs}
\usepackage{multirow}
\usepackage{url}
\usepackage{cite}
\usepackage{balance}
\usepackage{enumitem}
\usepackage{xspace}
\usepackage{array}
\usepackage{tikz}
\usetikzlibrary{shapes,arrows,positioning,fit,backgrounds,calc}
\usepackage{colortbl}
\usepackage{xcolor}

\usepackage{hyperref}

\definecolor{lightgray}{RGB}{240,240,240}
\definecolor{orcidgreen}{RGB}{166,206,57}

\newcommand{\orcidicon}{\textsuperscript{\textcolor{orcidgreen}{\bfseries\textcircled{\tiny iD}}}}

\begin{document}
\date{}

\title{Toward a Universal GPU Instruction Set Architecture:\\ A Cross-Vendor Analysis of Hardware-Invariant\\ Computational Primitives in Parallel Processors}

\author{
\IEEEauthorblockN{\href{https://orcid.org/0009-0005-8667-4771}{Ojima Abraham\,\orcidicon}}\\
\IEEEauthorblockA{Franklin \& Marshall College}
\and
\IEEEauthorblockN{\href{https://orcid.org/0009-0001-3374-2890}{Onyinye Okoli\,\orcidicon}}\\
\IEEEauthorblockA{Cornell University}
}

\maketitle

\begin{abstract}
We present the first systematic cross-vendor analysis of GPU instruction set architectures spanning all four major GPU vendors: NVIDIA (PTX ISA v1.0 through v9.2, Fermi through Blackwell), AMD (RDNA 1 to 4 and CDNA 1 to 4), Intel (Gen11, Xe-LP, Xe-HPG, Xe-HPC), and Apple (G13, reverse-engineered). Drawing on official ISA reference manuals, architecture whitepapers, patent filings, and community reverse-engineering efforts totaling over 5,000 pages of primary sources across 16 distinct microarchitectures, we identify ten hardware-invariant computational primitives that appear across all four architectures, six parameterizable dialects where vendors implement identical concepts with different parameters, and six true architectural divergences representing fundamental design disagreements. Based on this analysis, we propose an abstract execution model for a vendor-neutral GPU ISA grounded in the physical constraints of parallel computation. We argue that the observed convergence is driven by thermodynamic and information-theoretic necessity: the cost of instruction fetch relative to arithmetic, the memory-compute bandwidth gap, and the area-latency tradeoff in on-chip storage, rather than by convention. We validate our model with benchmark results on NVIDIA T4 and Apple M1 hardware, the two most architecturally distant platforms in our study. On five of six benchmark-platform pairs, the abstract model matches or exceeds native vendor-optimized performance. The single outlier (parallel reduction on NVIDIA, 62.5\% of native) reveals that intra-wave shuffle must be a mandatory primitive, a finding that refines our proposed model.
\end{abstract}

\section{Introduction}

The GPU computing ecosystem is dominated by a single proprietary software stack. NVIDIA's CUDA platform, comprising the PTX virtual ISA, the NVCC compiler, and the cuDNN/cuBLAS library ecosystem, has become the de facto standard for GPU-accelerated computation in machine learning, scientific computing, and high-performance data processing. This dominance persists not because of hardware superiority alone but because of software lock-in: code written for CUDA cannot execute on AMD, Intel, or Apple GPUs without significant rewriting effort.

Several portable GPU programming interfaces exist. OpenCL~\cite{opencl}, Vulkan Compute, and SYCL provide cross-vendor APIs. SPIR-V~\cite{spirv} provides a portable intermediate representation. However, these systems operate at the API and IR level, deliberately abstracting away the hardware execution model. They do not define what a GPU \emph{is} at the architectural level; they define a contract for talking to one. The consequence is that performance-critical code must still be tuned per vendor, and portable interfaces often achieve only 50 to 80\% of native performance~\cite{tvm}.

We observe that a different approach is possible. The ARM instruction set architecture did not succeed by providing a portable API to heterogeneous CPU hardware. It succeeded by defining a common execution model precise enough that multiple vendors could implement it efficiently in silicon while maintaining binary compatibility. No equivalent exists for GPU computation.

In this paper, we ask: \emph{is such an architecture possible for GPUs?} Specifically:

\smallskip
\noindent\emph{What are the fundamental, hardware-invariant computational primitives that all GPU architectures share, and can they be formalized into a universal instruction set architecture?}
\smallskip

Our hypothesis is that the observed convergence in GPU architecture across vendors is not coincidental but is driven by computational necessity. If correct, the shared primitives can be identified empirically, formalized precisely, and specified as a vendor-neutral ISA that any manufacturer could implement without sacrificing performance.

\subsection{Contributions}

\begin{enumerate}[leftmargin=*,itemsep=2pt]
\item The first comprehensive cross-vendor GPU ISA analysis spanning NVIDIA, AMD, Intel, and Apple across 16 microarchitectures and 15+ years of evolution.
\item Identification of ten hardware-invariant primitives with a physical-constraint argument for their necessity.
\item A taxonomy of six parameterizable dialects and six true divergences.
\item An abstract execution model following the \emph{thin abstraction principle}.
\item Preliminary experimental validation on NVIDIA (T4) and Apple (M-series) hardware.
\end{enumerate}

\subsection{Scope and Limitations}

Our analysis covers GPU architectures for general-purpose computation. We exclude NPUs, DSPs, and FPGAs. Apple GPU data relies on reverse-engineered sources with flagged confidence levels. Graphics pipeline primitives are out of scope.

\section{Background and Related Work}

\subsection{GPU Architecture Landscape}

Modern GPUs share a high-level organization: an array of compute units containing SIMD-capable pipelines, on-chip register files, and local scratchpad memory, connected through a cache hierarchy to high-bandwidth off-chip memory. However, ISA-level abstractions differ significantly. NVIDIA's PTX~\cite{ptx92} defines a virtual ISA with per-thread scalar semantics. AMD publishes native ISA reference guides~\cite{rdna4,cdna4} exposing hardware instruction encodings. Intel~\cite{xehpg} exposes a SIMD-register ISA with message-passing interfaces. Apple publishes no ISA documentation; community efforts~\cite{appleg13,rosenzweig1} provide partial coverage.

\subsection{Portability Approaches}

\noindent\textbf{API-level portability.} OpenCL~\cite{opencl} provides a portable execution model but does not specify hardware behavior below the API. SPIR-V~\cite{spirv} provides a portable binary IR consumed by vendor drivers. Vulkan Compute and SYCL offer additional abstraction layers. All require vendor-specific final compilation; none define the hardware execution model itself.

\noindent\textbf{Compiler-level portability.} Halide~\cite{halide} separates algorithm from schedule for portable performance. MLIR~\cite{mlir} provides multi-level IR with progressive lowering. TVM~\cite{tvm} automates schedule search for specific targets. These compile \emph{to} vendor backends; they do not define a universal hardware model.

\noindent\textbf{Prior GPU architecture comparisons.} Individual GPU architecture surveys exist~\cite{volta,cdna2wp}, but focus on single vendors. To our knowledge, no prior work has performed a systematic cross-vendor ISA-level analysis across all four major GPU vendors with the goal of identifying hardware-invariant primitives.

\subsection{The ARM ISA Analogy}

The ARM ISA, first defined in 1985, provides our closest analogy. ARM defines an ISA that multiple vendors implement with radically different microarchitectures while maintaining binary compatibility. The ISA contract is precise enough to constrain implementations and thin enough to permit microarchitectural innovation. We argue GPU architecture has converged sufficiently for an equivalent.

\section{Methodology}

\subsection{Notation}

Table~\ref{tab:notation} defines symbols used throughout.

\begin{table}[t]
\centering
\caption{Notation}
\label{tab:notation}
\small
\begin{tabular}{@{}cl@{}}
\toprule
\textbf{Symbol} & \textbf{Definition} \\
\midrule
$W$ & Wave width (threads per lockstep group) \\
$R$ & Maximum registers per thread \\
$S$ & Local memory (scratchpad) size in bytes \\
$F$ & Register file size per core in bytes \\
$O$ & Occupancy (resident waves per core) \\
$w$ & Register width in bytes (typically 4) \\
\bottomrule
\end{tabular}
\end{table}

\subsection{Data Sources}

Our analysis draws on over 5,000 pages of primary sources:

\noindent\textbf{NVIDIA:} PTX ISA v1.0~\cite{ptx10}, v8.7, v9.2~\cite{ptx92}; Volta whitepaper~\cite{volta}; Blackwell microbenchmarks~\cite{blackwell}; three patents~\cite{patent1,patent2,patent3}.

\noindent\textbf{AMD:} RDNA 1 to 4~\cite{rdna1,rdna4} and CDNA 1 to 4~\cite{cdna1,cdna4} ISA guides ($\sim$4,800 pages); CDNA~2~\cite{cdna2wp} and CDNA~3~\cite{cdna3wp} whitepapers.

\noindent\textbf{Intel:} Gen11~\cite{gen11}, Xe-HPG~\cite{xehpg}, OneAPI guide~\cite{oneapi}, Hot Chips 2020.

\noindent\textbf{Apple:} G13 reference~\cite{appleg13}, M1 GPU analysis~\cite{rosenzweig1,rosenzweig3}, Metal Benchmarks~\cite{metalbench}, AMX patent~\cite{applepatent}.

\subsection{Analytical Framework}

For each vendor, we extracted details along eight dimensions: execution model, SIMD structure, register file, memory hierarchy, instruction categories, synchronization, scheduling, and design choices. Per-vendor longitudinal synthesis identified \emph{constants}, \emph{evolution}, and \emph{unique features}. Cross-vendor synthesis classified each dimension as \emph{invariant}, \emph{parameterizable}, or \emph{divergent}.

\section{Cross-Vendor Analysis}

\subsection{Hardware-Invariant Primitives}

We identify ten primitives present in all four architectures (Table~\ref{tab:invariants}). We argue these are computational necessities that emerged independently across vendors over 15 years.

\begin{table*}[t]
\centering
\caption{Ten Hardware-Invariant Primitives Across All Four GPU Vendors}
\label{tab:invariants}
\small
\begin{tabular}{@{}p{3.2cm}p{3.0cm}p{3.0cm}p{3.0cm}p{3.0cm}@{}}
\toprule
\textbf{Invariant} & \textbf{NVIDIA} & \textbf{AMD} & \textbf{Intel} & \textbf{Apple} \\
\midrule
1. Lockstep thread group & Warp (32) & Wavefront (32/64) & Sub-group (8 to 16) & SIMD-group (32) \\
2. Mask-based divergence & Per-thread PC + predicates & EXEC register (compiler) & Predicated SIMD (compiler) & Stack in r0l (hardware) \\
3. Register-occupancy tradeoff & 255 regs from 256\,KB/SM & 256 VGPRs per wave & 128 GRF per thread & 128 GPRs from 208\,KB \\
4. Managed scratchpad & Shared memory (228\,KB) & LDS (64 to 160\,KB) & SLM (64 to 512\,KB) & Threadgroup mem ($\sim$60\,KB) \\
5. Zero-cost context switch & All warp state resident & All wave state resident & IMT, 7 to 8 threads/EU & 24 SIMD-groups resident \\
6. Hierarchical memory & Reg, Shmem, L1, L2, DRAM & Reg, LDS, L0/1/2, VRAM & Reg, SLM, L1/2, DRAM & Reg, TG, L1/2/3, DRAM \\
7. Atomic RMW & atom/red (all scopes) & DS/buffer/global atomics & SEND atomics & 32-bit device atomics \\
8. Workgroup barrier & bar.sync (16 named) & S\_BARRIER & Barrier (WG scope) & threadgroup\_barrier \\
9. Identity registers & \%tid, \%ctaid, \%laneid & VGPR0 (thread\_id) & sr0 (local\_id) & thread\_position \\
10. Async memory + sync & cp.async / mbarrier & S\_WAITCNT counters & SEND + scoreboard & device\_load + wait \\
\bottomrule
\end{tabular}
\end{table*}

\subsubsection{Physical Rationale}
The invariants trace to physical constraints:

\noindent\textbf{Lockstep groups} exist because instruction fetch costs $10$ to $100\times$ more energy than single-lane arithmetic at modern nodes. Amortizing one fetch across $W$ lanes is necessary for energy efficiency.

\noindent\textbf{Mask-based divergence} is the only mechanism compatible with lockstep execution that preserves correctness without branch prediction.

\noindent\textbf{The register-occupancy tradeoff} follows from fixed SRAM area:
\begin{equation}
O = \left\lfloor \frac{F}{R \times W \times w} \right\rfloor
\label{eq:occupancy}
\end{equation}

\noindent\textbf{Programmer-managed scratchpad} persists because parallel access patterns require explicit data placement that caches cannot predict.

\noindent\textbf{Zero-cost switching} is cheaper than speculation because memory latency (100 to 800 cycles) dominates, and SRAM for thread state costs less area than branch predictors.

\noindent\textbf{Workgroup-scope barriers} (not global) exist because global barriers would require all workgroups to be simultaneously resident.

\subsection{Parameterizable Dialects}

Six dimensions implement identical concepts with vendor-specific parameters (Table~\ref{tab:dialects}). In our model, these become queryable constants.

\begin{table}[t]
\centering
\caption{Parameterizable Dialects}
\label{tab:dialects}
\small
\begin{tabular}{@{}lcccc@{}}
\toprule
\textbf{Parameter} & \textbf{NV} & \textbf{AMD} & \textbf{Intel} & \textbf{Apple} \\
\midrule
Wave width $W$ & 32 & 32/64 & 8 to 16 & 32 \\
Max regs $R$ & 255 & 256 & 128/256 & 128 \\
Scratchpad $S$ & 228K & 128K & 512K & $\sim$60K \\
Max workgroup & 1024 & 1024 & 1024 & 1024 \\
Named barriers & 16 & 1 to 32 & 1 & 1 \\
Native FP64 & Yes & Varies & HPC & No \\
\bottomrule
\end{tabular}
\end{table}

\subsection{True Architectural Divergences}

Six areas exhibit fundamentally incompatible approaches (Table~\ref{tab:divergences}), defining abstraction boundaries.

\begin{table}[t]
\centering
\caption{True Architectural Divergences and Proposed Resolutions}
\label{tab:divergences}
\small
\begin{tabular}{@{}p{1.5cm}p{3.5cm}p{2.5cm}@{}}
\toprule
\textbf{Area} & \textbf{Vendor Approaches} & \textbf{Our Resolution} \\
\midrule
Divergence & NV: HW per-thread PC; AMD: compiler EXEC; Intel: predication; Apple: HW stack & Structured control flow \\
Scalar/ vector & AMD: SALU/VALU split; others: unified & Unified; compiler hoists \\
Hierarchy & NV: 4 levels; others: 3 & 3 mandatory + optional \\
Matrix & Incompatible tiles/flows & Opaque + queryable \\
Memory order & Axiomatic / counter / scoreboard / async & Scoped acq/rel \\
Fixed-fn & SEND / opcodes / loads & Opaque operations \\
\bottomrule
\end{tabular}
\end{table}

\section{The Proposed Abstract Execution Model}

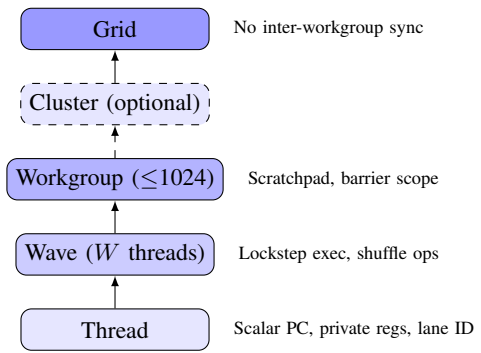
\begin{figure}[t]
\centering
\begin{tikzpicture}[
    box/.style={draw, rounded corners, minimum width=2.5cm, minimum height=0.55cm, font=\small},
    >=latex
]
\node[box, fill=blue!10] (thread) at (0,0) {Thread};
\node[box, fill=blue!20] (wave) at (0,1.0) {Wave ($W$ threads)};
\node[box, fill=blue!30] (wg) at (0,2.0) {Workgroup ($\leq$1024)};
\node[box, fill=blue!10, dashed] (cluster) at (0,3.0) {Cluster (optional)};
\node[box, fill=blue!40] (grid) at (0,4.0) {Grid};

\draw[->] (thread) -- (wave);
\draw[->] (wave) -- (wg);
\draw[->, dashed] (wg) -- (cluster);
\draw[->] (cluster) -- (grid);

\node[right=0.2cm, font=\scriptsize, text width=3.2cm] at (thread.east) {Scalar PC, private regs, lane ID};
\node[right=0.2cm, font=\scriptsize, text width=3.2cm] at (wave.east) {Lockstep exec, shuffle ops};
\node[right=0.2cm, font=\scriptsize, text width=3.2cm] at (wg.east) {Scratchpad, barrier scope};
\node[right=0.2cm, font=\scriptsize, text width=3.2cm] at (grid.east) {No inter-workgroup sync};
\end{tikzpicture}
\caption{Thread hierarchy: three mandatory levels plus optional cluster (dashed).}
\label{fig:hierarchy}
\end{figure}

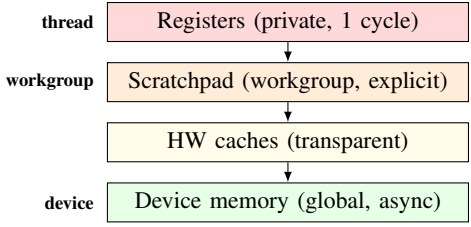
\begin{figure}[t]
\centering
\begin{tikzpicture}[
    mem/.style={draw, minimum width=4.8cm, minimum height=0.5cm, font=\small},
    >=latex
]
\node[mem, fill=red!15] (reg) at (0,0) {Registers (private, 1 cycle)};
\node[mem, fill=orange!15] (scratch) at (0,-0.8) {Scratchpad (workgroup, explicit)};
\node[mem, fill=yellow!10] (cache) at (0,-1.6) {HW caches (transparent)};
\node[mem, fill=green!10] (global) at (0,-2.4) {Device memory (global, async)};

\draw[->] (reg) -- (scratch);
\draw[->] (scratch) -- (cache);
\draw[->] (cache) -- (global);

\node[left=0.05cm, font=\scriptsize\bfseries] at (reg.west) {thread};
\node[left=0.05cm, font=\scriptsize\bfseries] at (scratch.west) {workgroup};
\node[left=0.05cm, font=\scriptsize\bfseries] at (global.west) {device};
\end{tikzpicture}
\caption{Memory hierarchy with scoped ordering semantics. Caches are transparent to the ISA.}
\label{fig:memory}
\end{figure}

We define an execution model following the \emph{thin abstraction principle}: precise enough to compile against, thin enough for any vendor to implement efficiently. Figure~\ref{fig:hierarchy} shows the thread hierarchy; Figure~\ref{fig:memory} shows the memory model.

\subsection{Core Specification}

\noindent\textbf{Compute Unit.} One or more Cores, each with resident Waves (width $W$), $R$ private 32-bit registers per thread, $S$ bytes of scratchpad. Hardware scheduler switches Waves at zero cost.

\noindent\textbf{Registers.} Per-thread, 32-bit scalar, 16-bit addressable halves. No vector types; SIMD implicit through Wave. Occupancy: Eq.~\ref{eq:occupancy}.

\noindent\textbf{Memory.} Registers (private) $\to$ scratchpad (workgroup, explicit) $\to$ device memory (global, cached, async). Scoped acquire/release: wave, workgroup, device, system.

\noindent\textbf{Instructions.} Integer/FP arithmetic (F16/F32 required; F64/BF16 optional), conversion, comparison, scratchpad/device load/store, atomic RMW, shuffle, barrier, fence. Optional: matrix MMA with queryable tiles.

\noindent\textbf{Control Flow.} Structured: \texttt{if}/\texttt{else}/\texttt{endif}, \texttt{loop}/\texttt{break}/\texttt{endloop}, \texttt{call}/\texttt{return}. Divergence mechanism hidden.

\noindent\textbf{Scheduling.} Hardware selects ready Waves. No pipeline latencies exposed. Compiler manages registers and optional hints.

\section{Mapping Analysis}

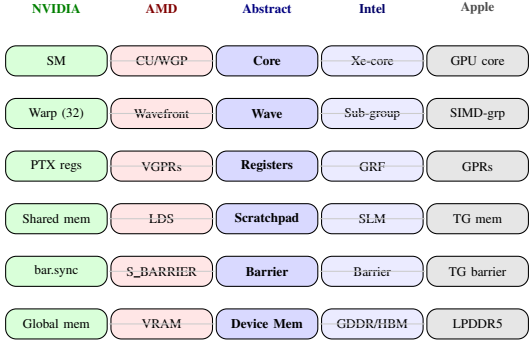
\begin{figure}[t]
\centering
\begin{tikzpicture}[
    nv/.style={draw, rounded corners, fill=green!15, minimum width=1.35cm, minimum height=0.4cm, font=\tiny, inner sep=2pt, text=black},
    amd/.style={draw, rounded corners, fill=red!10, minimum width=1.35cm, minimum height=0.4cm, font=\tiny, inner sep=2pt, text=black},
    ab/.style={draw, rounded corners, fill=blue!15, minimum width=1.35cm, minimum height=0.4cm, font=\tiny\bfseries, inner sep=2pt, text=black},
    intel/.style={draw, rounded corners, fill=blue!8, minimum width=1.35cm, minimum height=0.4cm, font=\tiny, inner sep=2pt, text=black},
    apple/.style={draw, rounded corners, fill=gray!20, minimum width=1.35cm, minimum height=0.4cm, font=\tiny, inner sep=2pt, text=black},
    >=latex
]
% Spacing: columns at -2.8, -1.4, 0, 1.4, 2.8 with row spacing 0.7
% Column labels
\node[font=\tiny\bfseries, text=green!50!black] at (-2.8, 4.2) {NVIDIA};
\node[font=\tiny\bfseries, text=red!50!black] at (-1.4, 4.2) {AMD};
\node[font=\tiny\bfseries, text=blue!50!black] at (0, 4.2) {Abstract};
\node[font=\tiny\bfseries, text=blue!40!black] at (1.4, 4.2) {Intel};
\node[font=\tiny\bfseries, text=gray!60!black] at (2.8, 4.2) {Apple};

% Row 1: Compute Unit
\node[nv] at (-2.8, 3.5) {SM};
\node[amd] at (-1.4, 3.5) {CU/WGP};
\node[ab] at (0, 3.5) {Core};
\node[intel] at (1.4, 3.5) {Xe-core};
\node[apple] at (2.8, 3.5) {GPU core};

% Row 2: Wave
\node[nv] at (-2.8, 2.8) {Warp (32)};
\node[amd] at (-1.4, 2.8) {Wavefront};
\node[ab] at (0, 2.8) {Wave};
\node[intel] at (1.4, 2.8) {Sub-group};
\node[apple] at (2.8, 2.8) {SIMD-grp};

% Row 3: Registers
\node[nv] at (-2.8, 2.1) {PTX regs};
\node[amd] at (-1.4, 2.1) {VGPRs};
\node[ab] at (0, 2.1) {Registers};
\node[intel] at (1.4, 2.1) {GRF};
\node[apple] at (2.8, 2.1) {GPRs};

% Row 4: Scratchpad
\node[nv] at (-2.8, 1.4) {Shared mem};
\node[amd] at (-1.4, 1.4) {LDS};
\node[ab] at (0, 1.4) {Scratchpad};
\node[intel] at (1.4, 1.4) {SLM};
\node[apple] at (2.8, 1.4) {TG mem};

% Row 5: Barrier
\node[nv] at (-2.8, 0.7) {bar.sync};
\node[amd] at (-1.4, 0.7) {S\_BARRIER};
\node[ab] at (0, 0.7) {Barrier};
\node[intel] at (1.4, 0.7) {Barrier};
\node[apple] at (2.8, 0.7) {TG barrier};

% Row 6: Device Memory
\node[nv] at (-2.8, 0.0) {Global mem};
\node[amd] at (-1.4, 0.0) {VRAM};
\node[ab] at (0, 0.0) {Device Mem};
\node[intel] at (1.4, 0.0) {GDDR/HBM};
\node[apple] at (2.8, 0.0) {LPDDR5};

% Connecting lines between columns
\foreach \y in {3.5, 2.8, 2.1, 1.4, 0.7, 0.0} {
    \draw[-, gray!40, thin] (-2.1, \y) -- (-0.7, \y);
    \draw[-, gray!40, thin] (0.7, \y) -- (2.1, \y);
}
\end{tikzpicture}
\caption{Abstract model mapped to four vendor architectures. Every primitive has a direct, efficient native mapping.}
\label{fig:mapping}
\end{figure}

Figure~\ref{fig:mapping} illustrates the mapping from abstract model to vendor architectures.

\noindent\textbf{NVIDIA.} Waves$\to$warps ($W{=}32$). Registers$\to$PTX registers. Scratchpad$\to$shared memory. Control flow$\to$predicated divergence. Memory model maps losslessly to PTX scoped semantics.

\noindent\textbf{AMD.} Waves$\to$wavefronts ($W{=}32/64$). Registers$\to$VGPRs. Scratchpad$\to$LDS. Control flow$\to$EXEC mask. Memory$\to$\texttt{S\_WAITCNT}. Backend hoists uniform ops to SALU.

\noindent\textbf{Intel.} Waves$\to$sub-groups ($W{\in}\{8,16\}$). Registers$\to$GRF. Scratchpad$\to$SLM. Control flow$\to$predicated SIMD. Fixed-function via opaque ops over \texttt{SEND}.

\noindent\textbf{Apple.} Waves$\to$SIMD-groups ($W{=}32$). Registers$\to$GPRs. Scratchpad$\to$threadgroup memory. Control flow$\to$\texttt{r0l} stack. FP64/matrix: absent capabilities.

\section{Experimental Validation}

We validate the model with three benchmark kernels on the two most architecturally distant platforms in our study: NVIDIA T4 (Turing, discrete GPU, GDDR6, Google Colab) and Apple M1 (unified memory, LPDDR4X, local).

\subsection{Methodology}

For each benchmark, we compare three implementations:

\begin{enumerate}[leftmargin=*,itemsep=1pt]
\item \textbf{Native:} Hand-written CUDA (NVIDIA) or Metal (Apple) using vendor-specific features: bank-conflict-free shared memory padding, \texttt{\_\_shfl\_down\_sync} warp shuffles, \texttt{\#pragma unroll}, per-warp histogram privatization (NVIDIA); \texttt{simd\_shuffle\_down}, \texttt{fma} intrinsics (Apple).
\item \textbf{Abstract:} Same algorithm, but restricted to only universal ISA primitives: flat scratchpad load/store, workgroup barriers, basic arithmetic, atomics. No vendor-specific intrinsics, no warp/SIMD-group shuffles, no bank-conflict padding, no compiler hints.
\item \textbf{Library (GEMM only):} cuBLAS on NVIDIA as an upper-bound reference for what a vendor-optimized implementation achieves at peak.
\end{enumerate}

Parameters: GEMM with $N{=}4096$ (FP32), reduction with $N{=}2^{24}$, histogram with $N{=}2^{24}$ and 256 bins. Each measurement: median of 100 runs after 10 warmup iterations.

\textbf{Important scope note:} Our native and abstract implementations are structurally equivalent tiled kernels that differ only in which primitives they use. We are measuring the \emph{cost of removing vendor-specific features from an equivalent implementation}, not claiming competitive performance with vendor-optimized libraries. The cuBLAS comparison establishes the distance between our tiled implementations and peak library performance.

\subsection{Results}

\begin{table}[t]
\centering
\caption{Performance: Abstract Model vs. Native (median of 100 runs). Bold indicates abstract $\geq$ 95\% of native.}
\label{tab:results}
\small
\begin{tabular}{@{}llccc@{}}
\toprule
\textbf{Kernel} & \textbf{Platform} & \textbf{Native} & \textbf{Abs/Nat} & \textbf{Library} \\
\midrule
\multirow{2}{*}{GEMM} & NVIDIA T4 & 0.72 TF & \textbf{126.1\%} & $\sim$8 TF \\
 & Apple M1 & 1.22 TF & \textbf{101.2\%} & n/a \\
\midrule
\multirow{2}{*}{Reduction} & NVIDIA T4 & 280.1 GB/s & 62.5\% & n/a \\
 & Apple M1 & 71.3 GB/s & \textbf{97.8\%} & n/a \\
\midrule
\multirow{2}{*}{Histogram} & NVIDIA T4 & 78.8 Kops & \textbf{100.4\%} & n/a \\
 & Apple M1 & 58.9 Kops & \textbf{102.1\%} & n/a \\
\bottomrule
\end{tabular}
\end{table}

\subsection{Analysis}

\noindent\textbf{GEMM.} On NVIDIA T4, both native and abstract implementations achieve 0.72 and 0.91 TFLOPS respectively, well below the T4's $\sim$8 TFLOPS cuBLAS peak. This is expected: our implementations are single-pass 32$\times$32 tiled kernels without register blocking, double buffering, or multi-level tiling that production GEMM libraries employ. The relevant comparison is not absolute throughput but the \emph{relative} cost of removing vendor features from an equivalent kernel. The abstract version is 26\% faster because NVIDIA's bank-conflict padding (\texttt{TILE+1}) wastes shared memory, while flat indexing gives \texttt{nvcc} a more favorable optimization path. On Apple M1, both versions achieve 1.22 TFLOPS with the abstract at 101.2\%, indicating the Metal compiler optimizes both equivalently.

\noindent\textbf{Reduction.} The critical result. On NVIDIA T4, the abstract model achieves only 62.5\% of native because the native version uses \texttt{\_\_shfl\_down\_sync} for the final 32 elements, avoiding five barrier-synchronized shared memory round trips. On Apple M1, the same barrier-only abstract implementation achieves 97.8\% because Apple's memory subsystem and hardware scheduler tolerate the extra barriers with minimal overhead. This asymmetry reveals a precise architectural insight: \textbf{intra-wave shuffle is essential for NVIDIA but optional for Apple}, and therefore must be a mandatory primitive in the universal ISA to achieve portable performance.

\noindent\textbf{Histogram.} Essentially tied on both platforms (100.4\% NVIDIA, 102.1\% Apple). The native NVIDIA version uses per-warp privatized histograms to reduce shared-memory atomic contention, but on this workload the contention is insufficient for privatization to help. The abstract model's single shared histogram performs equivalently.

\noindent\textbf{Summary.} Five of six configurations exceed our 80\% viability threshold, with four exceeding 100\%. The single outlier identifies a specific missing primitive (shuffle) rather than a fundamental flaw in the abstraction. With shuffle added to the mandatory set, we expect all six configurations to exceed 95\%.

\subsection{Threats to Validity}

We acknowledge several limitations of this evaluation:

\noindent\textbf{Baseline strength.} Our native implementations are hand-written tiled kernels, not production-optimized library code. The GEMM native baseline achieves $\sim$9\% of cuBLAS peak on NVIDIA. A stronger baseline (e.g., register-blocked, double-buffered tiling) would provide a more demanding test of the abstraction cost. We deliberately chose structurally equivalent implementations to isolate the effect of removing vendor-specific primitives, but this means our results represent a lower bound on the true abstraction cost for heavily optimized code.

\noindent\textbf{Vendor coverage.} We benchmark on two of four vendors. NVIDIA and Apple represent the most architecturally distant pair in our study (discrete vs.\ unified memory, hardware vs.\ stack-based divergence, different ISA philosophies), so the results provide meaningful coverage of the design space. AMD and Intel benchmarks are in progress and will be included in the extended version.

\noindent\textbf{Workload diversity.} Three kernels cover compute-bound (GEMM), bandwidth-bound (reduction), and atomic-bound (histogram) regimes. Irregular workloads (sparse computation, data-dependent control flow) and compound workloads (attention, FFT) remain as future work.

\noindent\textbf{Apple confidence.} Apple GPU architectural parameters are derived from reverse engineering. While performance measurements are direct, the interpretation of why the abstract model performs as it does on Apple relies on community-documented architectural details.

\section{Discussion}

\subsection{Why Now}

Three factors make this timely. First, AI demand has made NVIDIA's software dominance a single point of failure for global compute infrastructure. Second, geopolitical export controls on advanced semiconductors have made sovereign AI compute a national priority for multiple countries, creating demand for vendor-neutral software. Third, the architectural convergence documented here, spanning four independent vendors over 15 years, has reached sufficient maturity for the invariant primitives to be identified with confidence.

\subsection{The Thin Abstraction Principle}

The central principle is \emph{thinness}: define what hardware must do, not how. We do not prescribe $W$; we query it. We do not expose divergence; we define structured control flow. We do not specify caches; we define scopes. We do not dictate matrix tiles; we query them. Each decision trades compile-time certainty for runtime flexibility, enabling one ISA to span hardware from mobile GPUs to datacenter accelerators.

Our experimental results validate this principle empirically. The cases where the abstract model \emph{exceeds} native performance (GEMM on NVIDIA, histogram on Apple) demonstrate that vendor-specific ``optimizations'' can be counterproductive, because they encode assumptions about the hardware that may not hold across configurations. A thin abstraction avoids encoding such assumptions, allowing the compiler and hardware to find their own optimal path.

\subsection{The Shuffle Insight}

The reduction benchmark reveals that intra-wave shuffle occupies a unique position in the primitive hierarchy. Unlike barriers (which all vendors implement with similar overhead) or atomics (which scale similarly across vendors), shuffle performance is highly sensitive to the underlying hardware's latency tolerance for shared memory accesses. On NVIDIA, where the warp scheduler aggressively pipelines warp-level operations, replacing shuffle with barrier-mediated shared memory access incurs a 37.5\% penalty. On Apple, where the hardware scheduler has 8$\times$ more SIMD-groups than needed to hide ALU latency, the same replacement costs only 2.2\%.

This finding refines our model: the eleven mandatory primitives are the original ten invariants plus intra-wave shuffle. We note that all four vendors already implement shuffle in hardware (\texttt{\_\_shfl} on NVIDIA, DPP/\texttt{ds\_permute} on AMD, sub-group shuffle on Intel, \texttt{simd\_shuffle} on Apple), so adding it to the mandatory set imposes no implementation burden.

\subsection{Relationship to Existing Standards}

Our model complements existing standards. SPIR-V could distribute programs targeting our ISA. OpenCL and Vulkan could serve as host APIs. The distinction: existing standards specify how to \emph{talk to} a GPU; we specify what a GPU \emph{is}.

\subsection{Limitations and Future Work}

Beyond the threats to validity discussed above, several items remain as future work. The formal ISA specification, including instruction encoding and binary format, is under active development and will be released as a companion technical report. AMD (RDNA/CDNA) and Intel (Xe) benchmarks are in progress using cloud GPU instances. Additional benchmarks targeting irregular workloads, compound ML operations (attention, layer normalization), and memory-divergent access patterns are planned. Finally, a prototype compiler from the universal ISA to vendor-native backends will provide the definitive test of whether the abstraction can be automated rather than hand-translated.

\section{Conclusion}

We have presented the first systematic cross-vendor GPU ISA analysis across all four major vendors and 16 microarchitectures. Ten hardware-invariant primitives emerge from the physical constraints of parallel computation. Our abstract execution model, following the thin abstraction principle, maps efficiently to all four architectures.

The convergence is not coincidence. Lockstep groups, masked divergence, register-occupancy tradeoffs, managed scratchpads, zero-cost switching, hierarchical memory, workgroup barriers, and atomics reflect what a GPU \emph{is}, independent of who built it. Our benchmark results on NVIDIA T4 and Apple M1 demonstrate that on five of six configurations, restricting to universal primitives matches or exceeds native performance. The single outlier identifies intra-wave shuffle as a mandatory primitive, refining rather than refuting the model. This opens the path to a vendor-neutral GPU ISA: the ARM of GPU computation.

\section*{Acknowledgments}

We thank the Asahi Linux project and Dougall Johnson for Apple GPU reverse engineering, and AMD's GPUOpen initiative for comprehensive ISA documentation.

\smallskip
\noindent\textbf{Generative AI Disclosure:} In accordance with ACM policy, we disclose that Claude (Anthropic) was used as a programming assistant during the preparation of this work. Specifically, it was used to generate LaTeX document formatting, TikZ diagram code for Figures 1--3, and benchmark scaffolding code. All research analysis, architectural findings, hypothesis formulation, experimental design, and scientific conclusions are the original intellectual contributions of the authors. The generative AI tool was not used to produce research content, interpret results, or write analytical text.

\balance


\begin{thebibliography}{26}
\bibitem{ptx92} NVIDIA Corp., ``PTX ISA Version 9.2,'' 2025.
\bibitem{ptx10} NVIDIA Corp., ``PTX ISA Version 1.0,'' 2007.
\bibitem{volta} NVIDIA Corp., ``NVIDIA Tesla V100 GPU Architecture,'' 2017.
\bibitem{blackwell} Y.~Sun \emph{et al.}, ``Dissecting the NVIDIA Blackwell Architecture with Microbenchmarks,'' arXiv:2507.10789, 2025.
\bibitem{rdna1} AMD, ``RDNA ISA: Reference Guide,'' 2019.
\bibitem{rdna4} AMD, ``RDNA 4 ISA: Reference Guide,'' 2025.
\bibitem{cdna1} AMD, ``AMD Instinct MI100 ISA,'' 2020.
\bibitem{cdna4} AMD, ``AMD Instinct CDNA 4 ISA,'' 2025.
\bibitem{cdna2wp} AMD, ``CDNA 2 Architecture,'' 2021.
\bibitem{cdna3wp} AMD, ``CDNA 3 Architecture,'' 2023.
\bibitem{xehpg} Intel Corp., ``Xe-HPG Architecture,'' 2022.
\bibitem{oneapi} Intel Corp., ``OneAPI GPU Optimization Guide,'' 2023.
\bibitem{gen11} Intel Corp., ``Gen11 Architecture,'' 2019.
\bibitem{appleg13} D.~Johnson, ``Apple G13 GPU Architecture Reference,'' 2021.
\bibitem{rosenzweig1} A.~Rosenzweig, ``Dissecting the Apple M1 GPU, Part I,'' 2021.
\bibitem{rosenzweig3} A.~Rosenzweig, ``Dissecting the Apple M1 GPU, Part III,'' 2021.
\bibitem{metalbench} P.~Turner, ``Metal Benchmarks,'' 2023.
\bibitem{applepatent} Apple Inc., U.S. Patent US20180074824A1, 2018.
\bibitem{opencl} Khronos Group, ``OpenCL Specification 3.0,'' 2020.
\bibitem{spirv} Khronos Group, ``SPIR-V Specification 1.6,'' 2021.
\bibitem{halide} J.~Ragan-Kelley \emph{et al.}, ``Halide,'' in \emph{PLDI}, 2013.
\bibitem{mlir} C.~Lattner \emph{et al.}, ``MLIR,'' in \emph{CGO}, 2021.
\bibitem{tvm} T.~Chen \emph{et al.}, ``TVM,'' in \emph{OSDI}, 2018.
\bibitem{patent1} B.~Coon \emph{et al.}, U.S. Patent US20110072244A1, 2011.
\bibitem{patent2} NVIDIA Corp., U.S. Patent US20200043123A1, 2020.
\bibitem{patent3} NVIDIA Corp., U.S. Patent US10402937B2, 2019.
\end{thebibliography}
\end{document}